# Substrate engineering in the growth of perovskite crystals

Yu-Hao Deng[1*]

[1] Academy for Advanced Interdisciplinary Studies, Peking University, Beijing, China

* Correspondence should be addressed to yuhaodeng@pku.edu.cn

**Abstract**

Metal halide perovskites have recently emerged as promising materials for the next generation of optoelectronic devices owing to their remarkable intrinsic properties. In the growth of perovskite crystals, the substrates are essential and play a vital role. Herein, substrate engineering in the growth of perovskite crystals have been reviewed. Particularly, various modified strategies and corresponding mechanism based on the substrate engineering applied to the optimization of thickness, nucleation and growth rate are highlighted. Then the alterable adhesion to substrates will also be discussed. Furthermore, applying the structural coherence of epitaxial crystals with substrate, scalable perovskite single-crystalline thin films have been obtained and can be transferred onto arbitrary substrates. Substrate engineering also can stabilize the desired perovskite phases by modulating the strain between crystals and substrates. Finally, several key challenges and related solutions in the growth of perovskite crystals based on substrate engineering are proposed. This review aims to guide the future of substrate engineering in perovskite crystals for various optoelectronic applications.

**Introduction**

Metal halide perovskites, which can be synthesized via low-cost solution-based methods, have emerged as promising materials for the next generation of optoelectronic devices with intrinsic remarkable performance in various optoelectronic devices such as solar cells, light-emitting diodes (LED), lasers, photodetectors and X-ray imaging [1-8]. Substrates, which are essential for the preparation of materials and devices, have been shown to play a vital role in the growth of perovskite crystals. The surface energy of substrates can affect the heterogeneous nucleation of crystals [9]. Meanwhile, the interaction between substrate and precursor solution is critically important for ion diffusion in solvent, which determines the crystal growth rate and promotes the in-plane growth of the perovskite [7, 10-12]. In addition to the field of kinetics and thermodynamics, the adhesion between perovskite and substrate can also determine the quality of heterojunction interface and whether crystals can be transferred to fit the requirement of hybrid multilayer optoelectronic devices [13]. Moreover, when a crystal is compressed or stretched, the resulting deformation is called strain. For the ubiquitous strain, avoiding the negative impacts of strain and enabling strain be a highly effective tool for enhancing optoelectronic properties remain hot topics in the field of materials and devices [14]. Substrate engineering is so important in the growth of perovskite materials, but there is still no systematic summary in this field so far, thus it is urgent to review the knowledge about the applications of substrate engineering and how substrates affect crystals growth.

In this review, we highlight strategies for optimizing the thickness, nucleation, growth rate and adhesion to substrates of perovskite crystals to achieve better materials properties and device performance. Subsequently, applications of substrate engineering in coherent nano-, microstructure arrays and scalable single-crystalline thin films are discussed. Furthermore, controlling the strain between crystals and substrate and enhancing the phase stability of material by substrate engineering strain are also mentioned. Finally, several key issues and corresponding solutions in the growth of perovskite crystals based on substrate engineering are discussed. This review aims to provide guidelines of substrate engineering for the future design of perovskite crystals and for promoting their optoelectronic applications.

**1. Crystal thickness**

The performances of photoelectric devices are mainly determined by two factors: light absorption and carrier transfer, both of which are highly correlated with the crystal thickness [6]. Specifically, the excessively thick perovskite crystals would prolong the transit time of photon-generated carriers and increase the probability of non-radiative recombination, thus causing slower response and lower carrier collection efficiency. On the contrary, insufficient thickness would result in poor light absorption and degrade the energy conversion efficiency. Thus, thickness control of perovskites active layer is highly desirable for high-performance optoelectronic devices.

The space-confined method, restricting the precursor solution between two substrates and adjusting the thickness flexibly by altering the geometric space, is the direct application of substrate engineering in the field of perovskite growth [15]. As shown in Fig. 1A, a geometry-defined dynamic flow reaction system was built by separating two glass substrates with two spacers, to confine the crystal growth within the slit channel [16]. The thickness of the resulting MAPbI$_3$ crystals was controlled from 150 μm to 1440 μm by adjusting the gap between the two glass substrates. Yang et al. and Chen et al. applied pressure on the confined space respectively [6, 17]. Fig. 1B shows the customized mechanical device for pressure control between substrates. The thickness of the crystals decreases linearly with pressure under logarithmic coordinates over three orders and can be reduced to 100 nm level under $10^8$ Pa (Fig. 1C), which demonstrates the decisive role of pressure in thickness control.

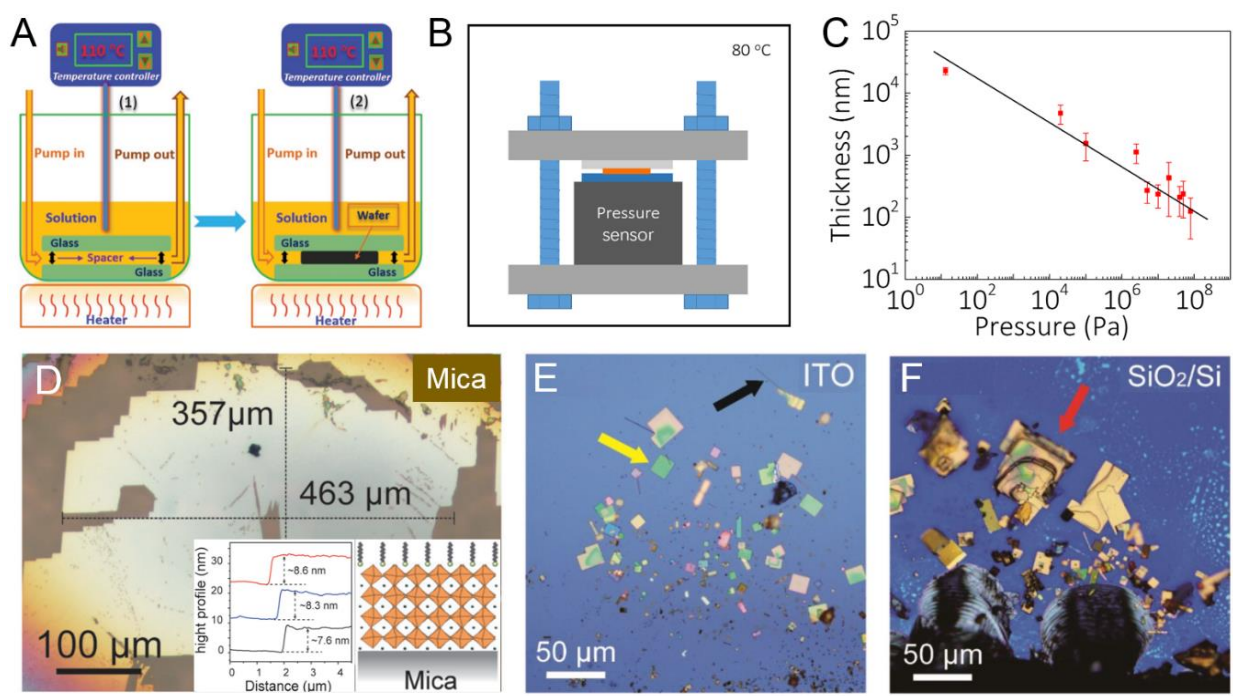

**Fig. 1.** Thickness adjustment of perovskite crystals. (A) Schematic diagram of space-confined method. (B) The schematic diagram of pressure control device. (C) The thickness of the film varies with different pressure growth conditions. (D) 2D growth of MAPbBr$_3$ perovskite on a mica substrate. (E) Perovskites grown on the ITO substrate. (F) Perovskites grown on the SiO$_2$/Si substrate. (A) Reproduced with permission from Ref. [16], ©WILEY-VCH Verlag GmbH & Co. KGaA, Weinheim 2016. (B, C) Reproduced with permission from Ref. [6], ©WILEY-VCH Verlag GmbH & Co. KGaA, Weinheim 2018. (D-F) Reproduced with permission from Ref. [12], © The Royal Society of Chemistry 2020.

Thickness control can also be achieved via the interactions between perovskite and substrate. Strong interactions can decrease the interface energy, driving the striking in-plane growth of the perovskite [12]. As shown in Fig. 1D, the strong potassium-halogen ionic interactions at the perovskite/mica interface provide a new kinetic route for controlling perovskite to 2D growth. The grown MAPbBr$_3$ crystal was characterized as 8 nm in thickness and hundreds of micrometers in lateral size. In contrast to mica, the grown MAPbBr$_3$ crystals on ITO, silicon and quartz substrates preferred to crystallize in solution rather than in-plane growth (Fig. 1E, F) [12].

By controlling the gap between substrates in space-confined method and the interactions between perovskite and substrate in epitaxy growth, substrate engineering has been applied in adjusting crystal thickness and providing an optimized thickness for high-performance optoelectronic devices.

## 2. Crystal nucleation

Generally, crystal growth contains two processes: nucleation and subsequent growth. High nucleation density limits the growth space for initial nucleus and induces the emergence of grain boundaries, which lead to low carrier mobility and high trap density and then degrade the device performances [10]. Therefore, inhibiting nucleation and providing enough space for nucleus growing up to enlarge the crystal size have been an urgent problem in the field of perovskite growth.

According to classical heterogeneous nucleation theory [9]. The free energy barrier needed for heterogeneous nucleation ($\Delta G_{het}$) is equal to the product of homogeneous nucleation free energy barrier ($\Delta G_{heom}$) and a function of the contact angle ($\theta$): $\Delta G_{het} = \Delta G_{heom} f(\theta)$, where $f(\theta) = \frac{2-3\cos\theta+\cos^3\theta}{4}$. For a particular solution system, the $\Delta G_{heom}$ is a constant value and independent of the substrate. The wettability of the substrate surface has a strong effect on the nucleation rate ($J$) via the inverse exponential dependence on $\Delta G_{het}$,

$$J = J_0\, e^{-\Delta G_{het}/kT} = J_0\, e^{-\Delta G_{heom}(2-3\cos\theta+\cos^3\theta)/4kT} \qquad (1)$$

where $J_0$ is a kinetic constant, $k$ is the Boltzmann constant and $T$ is the temperature. The numerical simulation shows in Fig. 2A. It can be seen that the increasing of contact angle (or substrate hydrophobicity) will lead to the decrease of the nucleation rate on substrate.

Deng et al. explained the mechanism of the influence of surface hydrophobicity on heterogeneous nucleation (Fig. 2B, C) [7]. Hydrophilic surface has a stronger attraction between surface atoms with solvent molecules and ions than hydrophobic surface [7, 18, 19]. The strong attraction will fix more pre-existing clusters on the substrate surface and block the re-dissolution of ions from the clusters, resulting in more stable existing clusters and easier to grow into crystal nucleus. Therefore, the crystal nucleation free energy barrier on hydrophilic substrate is relatively lower than the hydrophobic substrate, yielding a higher nucleation rate of crystals [9, 20]. The nucleation density of crystals decreases by 300 times from the completely hydrophilic to hydrophobic substrate surface (Fig. 2D, E). As the contact angle of the substrate increases, the measured nucleus densities on the substrates with different solvent contact angles shows in Fig. 2F.

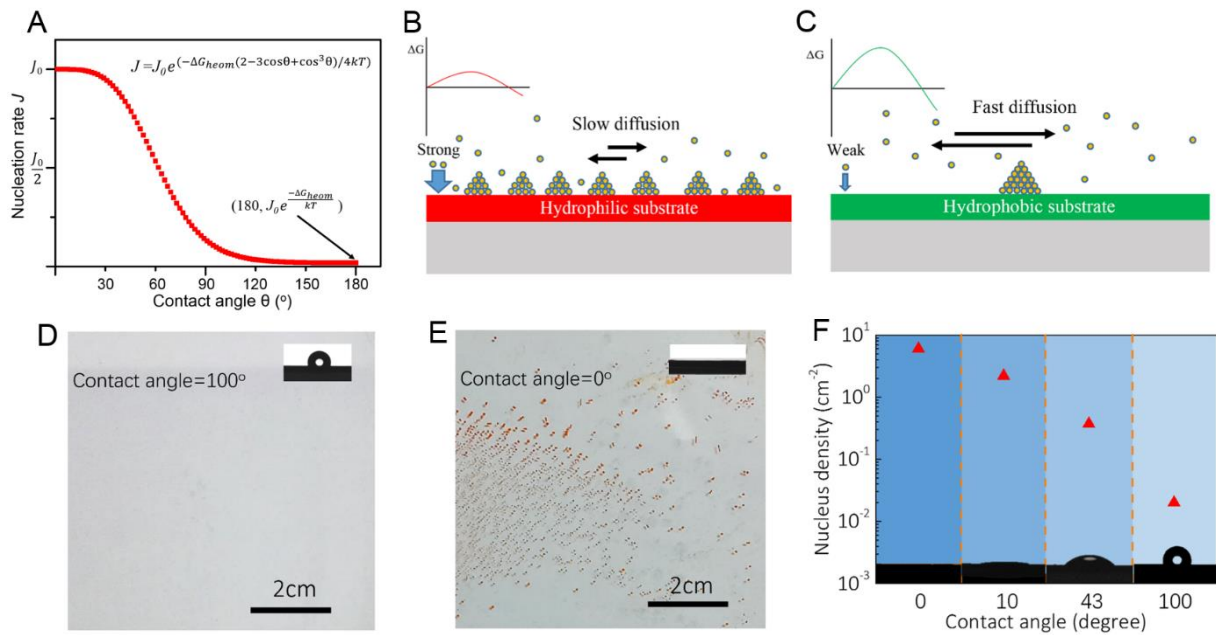

**Fig. 2.** Nucleation control of perovskite crystals. (A) Numerical simulation of equation (1). Schematic illustrations of the nucleus mechanism on hydrophilic (B) and hydrophobic (C) substrates. Inset: the energy barrier of forming nucleus. Photographs of nucleation on the hydrophobic substrate (D) and hydrophilic substrate (E). (F) The measured nucleus densities on substrates with different solvent contact angles. (B-F) Reproduced with permission from Ref. [7], © Springer Nature 2020.

Substrate engineering is a facile and effective way to control the crystal nucleation. The nucleation density of substrates can be adjusted flexibly via selective surface treatment, so as to achieve desirable crystal morphology on the substrate. By introducing the hydrophobic treatment on substrates to inhibit nucleation, Deng et al. successfully grown the single‑crystalline perovskite thin‑film with an aspect ratio of 1000 (1 cm in side length, 10 μm in thickness) under space-confined condition [7].

## 3. Crystal growth rate

Compared with perovskite polycrystalline films, grain-boundary-free perovskite single-crystalline thin film possesses lower trap-state density, higher carrier mobility, longer diffusion length and higher stability, which are supposed to maximize the performance of solar cells and other optoelectronic devices [21]. For most applications, large-area single-crystalline thin films are required to reduce the cost of market application. However, the crystal nucleation on the substrate is random and thus new nucleus would inevitably form during the long growth process. If the growth rate can be extremely high, the nucleus would grow up so rapidly that new nucleus does not form. Therefore, the faster growth is ideal to synthesize large-area single-crystal perovskite films.

When the space is down to micrometer scale, the interaction between substrate surface and precursor solution becomes critically important for ion diffusion in solvent and determines how fast the crystal growth rate. As illustrated in Fig. 3A, B [18], the large surface tension on wetting substrates imposes a friction force that drags down the ion diffusion, which slows the diffusion of ions (Fig. 3C), leading to a slower crystal growth rate. Deng et al. reported the growth rate of the perovskite crystals increases with the increase of the contact angle (Fig. 3D) [7]. The average growth rate of crystals in the gap of hydrophobic substrates is 31.5 μm/min, which is 6-fold faster than that on the completely hydrophilic substrates. Notably, the growth rate of crystals in the micrometer-scale space between hydrophobic substrates is even higher than the growth rate of the bulk crystal in an absolutely free condition in solution (27.0 μm/min) [15].

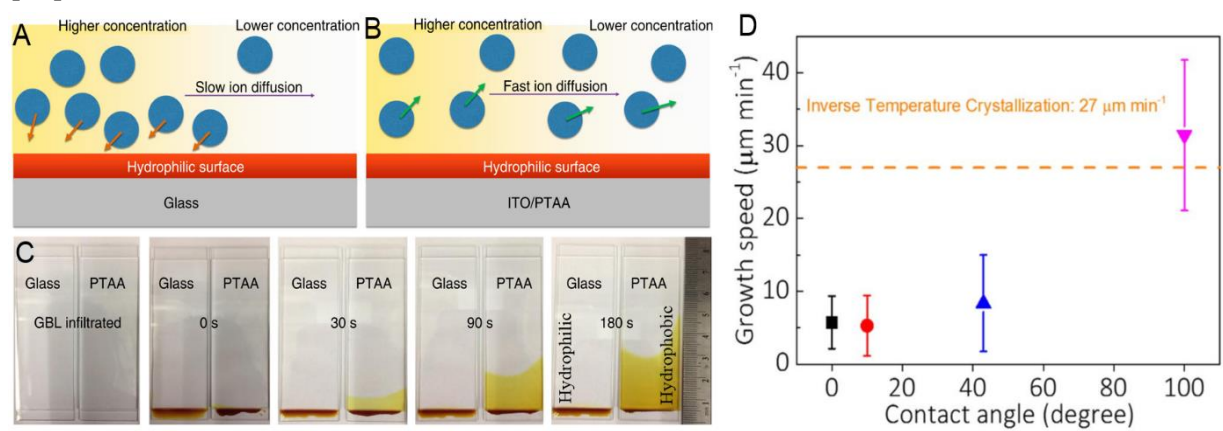

**Fig. 3.** Improving growth rates of perovskite crystals. Schematic illustrations of ion diffusion rate in the confined gaps using hydrophilic (A) and hydrophobic (B) substrates. (C) Photographs of the diffusion process of MAPbI$_3$ precursor solution in the confined gaps of hydrophilic glasses and hydrophobic substrates after different durations. (D) Growth rates of perovskite crystals on the substrates with different solvent contact angles. (A-C) Reproduced with permission from Ref. [18], © Springer Nature 2017. (D) Reproduced with permission from Ref. [7], © Springer Nature 2020.

Substrate engineering accelerates the growth rate of perovskite crystals by enhancing the diffusion rate of precursor ions. The fast growth of perovskite crystals opens an avenue for the production of large-scale single-crystalline thin films in short timescales, which is very important for low-cost and scalable perovskite industrial applications.

## 4. Adhesion to substrate

A high-quality interface of the perovskite/substrate heterojunction is promising to provide high-speed, high-efficiency and long-term stability performances for integrated optoelectronic devices [22]. After crystallization growth, the perovskite crystals are normally fixed on the given substrates with strong bonds, which also hinders perovskite to integrate with other functional semiconducting and metallic materials for hybrid multilayer optoelectronic devices. Substrate engineering can provide perovskite alterable adhesions with the substrate and open a new avenue for various application requirements.

Yang et al. formed stronger bonds in MAPbBr$_3$/ITO heterojunction by activating the dangling bond on ITO surface [6]. Wei et al. modified the Si substrate with a layer of brominated (3-aminopropyl) triethoxysilane (APTES) to connect the perovskite via primary chemical bonds [23]. Due to the chemical bonds formed at the interface, the perovskite crystals

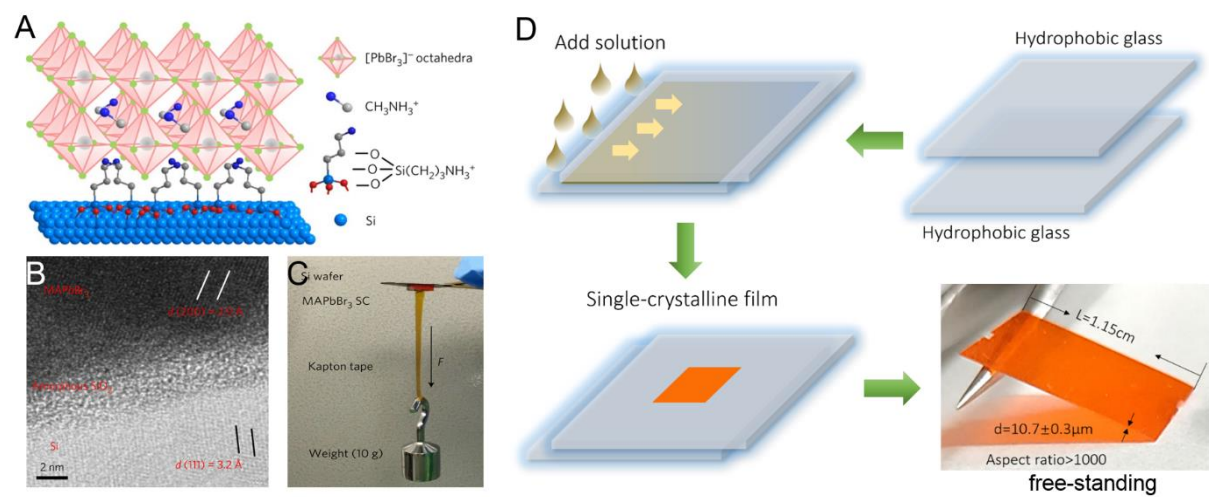

**Fig. 4.** Adhesion modulation of perovskite crystals. (A) Schematic illustration of the structure of Si-integrated MAPbBr$_3$ single crystals. (B) High-resolution TEM image of the cross-section of the interface of the Si-integrated MAPbBr$_3$ single crystal. (C) Photograph of Si-integrated MAPbBr$_3$ single crystal with a 10 g weight. (D) The schematic diagram of the growth process

between hydrophobic substrates and the free-standing single-crystalline thin film peeled off from the substrate. (A-C) Reproduced with permission from Ref. [23], © Springer Nature 2017. (D) Reproduced with permission from Ref. [7], © Springer Nature 2020.

grown on the functionalized Si wafers had a solid mechanical and electrical connection, which can bear a tensile pressure of 3.5 kPa (Fig. 4A). Relying on the hydrophobic treatment of the substrate to reduce surface adhesion, Ding et al. transferred millimeter-sized perovskite single-crystalline films onto other target substrates [13] and Deng et al. peeled off and transferred the centimeter-scale perovskite single-crystalline thin film from the substrate (Fig. 4B) [7].

The adhesion to the substrate can be adjusted by specific modification of the substrate. Substrate engineering broadens extensive applications of perovskite material for integrated optoelectronic systems and more complex multilayer devices.

## 5. Crystal orientation

Grain boundaries and unordered grain orientations in perovskite polycrystalline films result in high trap-state density, low carrier mobility and short diffusion length, which limit the potential performance of thin-film devices [24]. Perovskite single crystals with structural coherence on the atomic scale have been reported to exhibit superior properties to polycrystalline thin films. However, large-scale growth of single-crystalline thin films still suffer from the barriers of existing growth technologies such as random nucleation and limited growth rate [7]. The substrate-dependent epitaxial growth would offer a scalable, inexpensive and readily accessible route to coherent and continuous films that exhibit superior electronic and optical properties owing to the absence of high-angle grain boundaries.

As shown in Fig. 5A, Meagan et al. successfully grown the single crystal-like $CsPbBr_3$ films on a wafer-sized $SrTiO_3$(100) substrate by spin coating epitaxial method [25]. Highly ordered $MAPbBr_3$ on PbS, $CsPbBr_3$ on $SrTiO_3$ and $CsPbBr_3$ on GaN were also grown via liquid phase or gas phase epitaxy (Fig. 5B-D) [26-28]. Recently, Lei et al. reported a solution-based lithography-assisted epitaxial-growth-and-transfer method for fabricating single crystal hybrid perovskites on arbitrary substrates. The grown perovskite crystals possess precise control of their thickness (from about 600 nm to about 100 μm) and area (continuous thin films up to about 5.5 cm×5.5 cm) (Fig. 5E, F) [29].

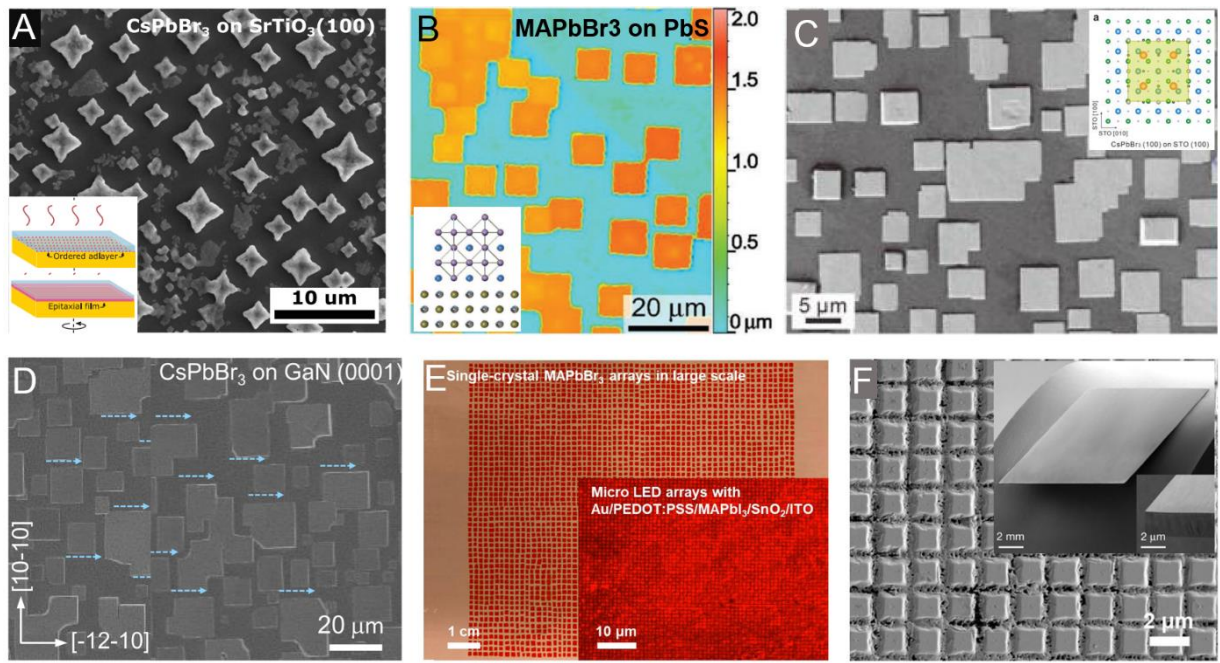

**Fig. 5.** Coherent crystal orientation of perovskite crystals. Epitaxial growth of CsPbBr$_3$ films on SrTiO$_3$ (100) substrate (A), MAPbBr$_3$ on PbS (B), CsPbBr$_3$ on SrTiO$_3$ (C) and CsPbBr$_3$ on GaN (D). Solution-based lithography-assisted epitaxial-growth method for fabricating scalable single crystal MAPbBr$_3$ perovskite (E) and MAPbI$_3$ perovskite on arbitrary substrates (F). (A) Reproduced with permission from Ref. [25], © American Association for the Advancement of Science 2019. (B) Reproduced with permission from Ref. [26], ©WILEY-VCH Verlag GmbH & Co. KGaA, Weinheim 2020. (C) Reproduced with permission from Ref. [27], © American Chemical Society 2017. (D) Reproduced with permission from Ref. [28], © American Chemical Society 2019. (E, F) Reproduced with permission from Ref. [29], © Springer Nature 2020.

By enabling the continuation of the substrates atomic lattice by the perovskite epitaxial layer, coherent nano-, microstructure arrays and scalable single-crystalline thin films have been obtained on selected substrates. Substrate engineering provides a new platform to advance large-scale material growth and optoelectronics of perovskite single-crystalline thin films.

## 6. Strain and phase stability

The uncontrollable strain caused by mismatched thermal expansion of perovskite crystals and substrates during the thermal annealing process accelerates degradation of perovskite films under illumination [15]. Substrate engineering is a powerful tool to control the strain between crystals and substrate and even can enhance the phase stability of material by strain. For example, single-cation α-FA/CsPbI$_3$ materials possess the improved environmental stability, higher carrier mobility and stronger light absorption, has been the most potential application of halide perovskites in photovoltaics (solar cells) [30-32]. However, a major obstacle of the black α-phase is the tendency to transform into yellow hexagonal structure (δ-phase) at photovoltaic devices operating temperature.

As shown in Fig. 6A, the strain between the epitaxial crystals and substrate can constrain the lattice from the phase transition [33]. Chen et al. reported that the structure of optically active α-FAPbI$_3$ can be stabilized for at least a year at room temperature by growing it on a lattice-mismatched perovskite substrate (Fig. 6B, C) [34]. The strain also reduces the bandgap and increases the carrier mobility of α-FAPbI$_3$. The same concept of strain induced stabilization, Steele et al. applied the interfacial clamping and strain to form an RT-stable black phase of functional CsPbI$_3$-based thin films on ITO substrate (Fig. 6D, E) [35].

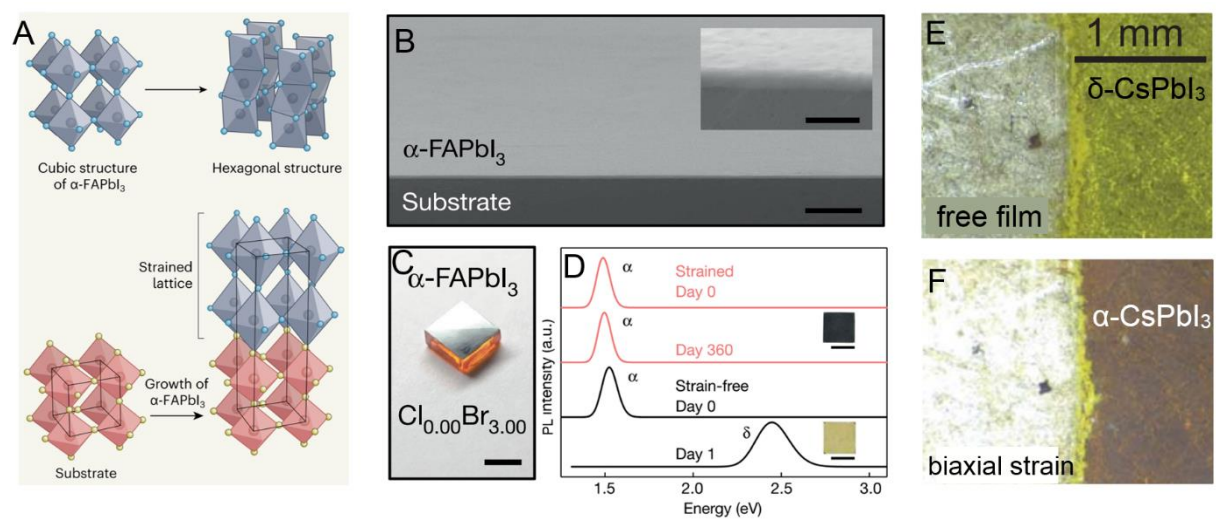

**Fig. 6.** Strain enhances the phase stability of perovskite crystals. (A) The schematic diagram of α-FAPbI$_3$ that is stabilized by growing it on a lattice-mismatched perovskite substrate. (B) A cross-sectional scanning electron microscope (SEM) image of the epitaxial thin film with controlled uniform thickness. (C) Optical image of the as-grown epitaxial α-FAPbI$_3$ thin film. (D) Phase stability study by photoluminescence spectroscopy for at least a year at room temperature. (E) Free yellow δ-phase CsPbI$_3$. (F) Black α-phase CsPbI$_3$ under biaxial strain. (A) Reproduced with permission from Ref. [33], © Springer Nature 2020. (B-D) Reproduced with permission from Ref. [34], © Springer Nature 2020. (E, F) Reproduced with permission from Ref. [35], © American Association for the Advancement of Science 2019.

Materials with high phase stability are the aiming goals in the research field of perovskite optoelectronic devices. Owing to the substantial stabilization effect in strain modulation, substrate engineering provides a new platform to favor the formation of a desired stable phase or can even lead to new phases.

## 7. Conclusions and perspective

In summary, substrate engineering is a facile and effective way for the controllable growth of perovskite crystals. By applying appropriate modified strategies, various widely concerned issues of perovskite crystal growth can be effectively alleviated, including thickness control, nucleation inhibition, growth rate, size enlargement, tunable adhesion, coherent orientation and phase stability. These improvements satisfy a majority of application requirements of perovskite optoelectronic devices well. Even though lots of achievements have been obtained by the substrate engineering, there are still some attentions that need to be addressed. For example, the growth of large-scale (>1 cm) ultra-thin (<400 nm) perovskite single-crystalline thin films via substrate engineering still remains the enormous challenge. Substrate morphology ought to play an influential role in the crystal growth, but the research and exploration in this area are insufficient [36]. Moreover, substrate engineering cannot solve the natural instability of organic-inorganic hybrid perovskite, all-inorganic or inorganic doping cation structure would provide a solution for the stability [37]. Finally, considering toxicity of the lead element, environmental protection and laboratory safety, lead-free perovskite, lead sequestration substrate and health protection should be realized by the further approaches [38-41]. All in all, applying the substrate engineering to the growth of perovskite crystals, scalable single-crystalline thin films, functionalized heterojunction interface, transferability and phase stability have been significantly improved. We believe that substrate engineering will be a fundamental driving force to enable perovskite an excellent alternative material for photovoltaic (solar cells) and other photoelectric applications in the coming future.

**Acknowledgements:** None.

**Conflict of interest:** The authors declare no competing financial interest.